\documentclass[12pt,reprint,aps,groupedaddress,showpacs,nofootinbib,prl]{revtex4-1}
\usepackage{amsmath,amssymb,graphicx,amsfonts}
\usepackage{mathrsfs}
\usepackage{xcolor}

\newcommand{\orr}{\omega_{r}}
\newcommand{\ot}{\omega_{\theta}}

\usepackage{hyperref}
\usepackage{setspace}

\begin{document}
\title{Gravitational-wave glitches in chaotic extreme-mass-ratio inspirals}

\author{Kyriakos Destounis}
\email{kyriakos.destounis@uni-tuebingen.de}
\affiliation{Theoretical Astrophysics, IAAT, University of T{\"u}bingen, Germany}

\author{Arthur G. Suvorov}
\affiliation{Theoretical Astrophysics, IAAT, University of T{\"u}bingen, Germany}

\author{Kostas D. Kokkotas}
\affiliation{Theoretical Astrophysics, IAAT, University of T{\"u}bingen, Germany}

\date{\today}
 
\begin{abstract}
\noindent{The Kerr geometry admits the Carter symmetry, which ensures that the geodesic equations are integrable. It is shown that gravitational waveforms associated with extreme-mass-ratio inspirals involving a non-integrable compact object display `glitch' phenomena, where the frequencies of gravitational waves increase abruptly, when the orbit crosses certain spacetime regions known as Birkhoff islands. The presence or absence of these features in data from upcoming space-borne detectors will therefore allow not only for tests of general relativity, but also of fundamental spacetime symmetries.}
\end{abstract}

\maketitle

\noindent{\bf\emph{Introduction.}} One of the main targets for space-borne gravitational-wave (GW) detectors, such as the Laser Interferometer Space Antenna (LISA) and Taiji \cite{lisafreq,taiji}, are extreme-mass-ratio inspirals (EMRIs). These binaries involve a $\text{(super-)massive}$ black hole (BH) and a companion whose mass is negligible (mass ratio $\lesssim 10^{-4}$) relative to the primary. This latter aspect, which implies that the inspiral dynamics are well approximated as the geodesic motion of the companion through the gravitational field generated by the primary, even close to the final plunge \cite{ori00} (though see also Ref. \cite{berry16}), means that EMRIs provide unparalleled information about the governing theory of gravity \cite{ryan95,glamp06,bara20}. In particular, general relativity (GR) predicts that the Kerr metric uniquely describes the gravitational field of astrophysically stable BHs \cite{rob75}. Therefore, if EMRIs with significantly non-Kerr trajectories present themselves in the LISA or Taiji data, this would represent a `smoking gun' for the breakdown of GR \cite{bert05,Barack:2006pq,schutz09}. 

One interesting possibility of a non-GR signature relates to the integrability of the orbit \cite{apo09,luk,cont11,luk14,glamp19,dest20}. The Kerr metric is stationary and axisymmetric and therefore possesses two Killing vectors, which lead to the conservation of energy and angular momentum for relativistic particles traversing the gravitational field. Additionally, the Kerr metric is sufficiently special, algebraically speaking, so as to admit a non-trivial Killing tensor \cite{carter68}, which provides an additional constant of motion and leads to the Liouville-integrability of the geodesic equations; Kerr geodesics do not display chaotic phenomena \cite{wilkins72}. Distortions in the Hamiltonian describing geodesic motion in the Kerr spacetime may either preserve the integrability or not depending on their exact character, in accord with the Kolmogorov-Arnold-Moser (KAM) theorem \cite{moser62,arnold63}. 

Even when a non-integrable perturbation is introduced into the Kerr Hamiltonian, some orbits necessarily remain periodic \cite{poin12,birk13}. From a phase-space perspective however, when the perturbation is non-integrable, small `islands' of stability come to surround the surviving periodic orbits \cite{arnold89}. These islands have the property that, as a particle passes through them, the ratio of longitudinal and transverse frequencies associated with the orbit remain constant \cite{Brink:2013nna,brink13}. In this way, the orbital dynamics are expected to display a transient plateau feature \cite{cont02} as gradual radiation reaction can cause orbits to weave in and out of islands. While this leads to clear signatures from a dynamical systems perspective \cite{apo09,luk,cont11,luk14,dest20}, we show here that occupancy in an island leads to a noticeable `glitch' (i.e., a rapid though short-lived increase, akin to that which is observed in pulsars) in the GW frequency. However, as observed in the LISA pathfinder mission \cite{path1}, instrumental noise artifacts can also lead to similar phenomena in the data stream (see Fig. 2 in Ref. \cite{ed20}). The main contribution of this Letter is to demonstrate that abrupt frequency increases during an EMRI may have a genuine astrophysical origin related to the fundamental spacetime symmetries associated with supermassive compact objects, and discarding them \emph{a priori} as instrumental artifacts may miss crucial physics.

\noindent{\bf\emph{Spacetime geometry.}} Theoretical deviations from a Kerr description for dark, compact objects can be broadly classified into two categories: those which introduce generic, though integrable, deformations into the geodesic Hamiltonian, and those which introduce non-integrable deformations \cite{vig11,joh13,krz16}. To give a concrete example of how these features appear at the level of the spacetime metric, we consider the geometry introduced in Ref. \cite{dest20}. In Boyer-Lindquist coordinates $(t,r,\theta,\phi)$, the metric we operate with reads
\begin{equation} \label{eq:genmet}
\begin{aligned}
ds^2 =&  -\frac{\Sigma[\left(\alpha_{Q}/r\right)M^3  + \Delta-a^2 A(r)^2 \sin^2\theta]}{[(r^2+a^2)-a^2 A(r) \sin^2\theta]^2} dt^2  \\
&- \frac{2 a [(r^2+a^2)A(r)-\Delta] \Sigma \sin^2\theta}{[(r^2+a^2)-a^2 A(r)\sin^2\theta]^2} dt d \phi \\
&+ \frac{\left(\alpha_{Q}/r\right)M^3  + \Sigma}{\Delta} dr^2 +\Sigma d \theta^2 \\
&+  \frac{\Sigma \sin^2 \theta \left[(r^2 + a^2)^2 - a^2 \Delta \sin^2 \theta \right]}{[(r^2+a^2) -a^2 A(r) \sin^2 \theta]^2} d \phi^2,
\end{aligned}
\end{equation} 
where $\Sigma = r^2 + a^2 \cos^2 \theta$, $\Delta = r^2 - 2 M r + a^2$, and $A(r) = 1 + r^{-2}  \alpha_{22} M^2$ for mass $M$ and spin $a$. The line element \eqref{eq:genmet} resembles the Kerr geometry, though contains two additional parameters. One of these, $\alpha_{22}$, was first considered by Johannsen \cite{joh13}, and introduces a generic, though integrable, deformation from Kerr (henceforth referred to as a `deformed-Kerr' metric). The other, $\alpha_{Q}$, breaks the Carter symmetry and leads to a non-integrable perturbation (`non-Kerr' metric) \cite{dest20}. In the limit where these two latter parameters vanish, the Kerr metric is recovered and the spacetime is an exact solution in GR and various other theories of gravity \cite{psal08}. For a wide range of values of $\alpha_{22}$ and $\alpha_{Q}$, the metric described by \eqref{eq:genmet} exists as an exact, vacuum solution in some particular family of mixed scalar-$f(R)$ theories \cite{suv20}. 

\noindent{\bf\emph{Geodesic dynamics and non-integrability.}} Noting that the metric components $g_{\mu \nu}$ within \eqref{eq:genmet} are functions of $r$ and $\theta$ only, so that the spacetime is stationary and axisymmetric, Hamilton's equations for a particle of mass $\mu$ with momenta $p^{\mu} = \dot{x}^{\mu}$ imply that $\dot{p}_{t} = 0 = \dot{p}_{\phi}$, where the overhead dot denotes differentiation with respect to proper time. In general, there exist three constants of motion associated with geodesic dynamics: the orbital energy, $E$, angular momentum, $L_{z}$, and the particle mass itself, $\mu$. The Kerr spacetime $(\alpha_{22} = \alpha_{Q} = 0$), in addition to admitting the three linear symmetries detailed above, also possesses a quadratic symmetry due to the existence of a non-trivial rank-2 Killing tensor \cite{carter68}. This additional integral of motion, giving rise to the Carter constant (which generally exists even for $\alpha_{Q} = 0,\,\alpha_{22} \neq 0$), implies the integrability of the geodesic equations as a whole \cite{cont02}.

A generic orbit for a stationary spacetime can be characterized by two libration-like frequencies, $\orr$ and $\ot$, which describe the transition rate between periastron and apastron and longitudinal oscillations around a given plane, respectively \cite{Brink:2013nna,brink13}. Those orbits for which $\orr$ is an integer multiple of $\ot$, or vice versa, are called resonant. At the level of orbital dynamics, the KAM and Poincar{\'e}-Birkhoff theorems \cite{moser62,arnold63} together imply that half of the resonant orbits remain stable while half become unstable when introducing a non-integrable `perturbation' into the Kerr (or any other regular) Hamiltonian. In the phase space, small islands of stability (`Birkhoff islands') come to surround each of the stable orbits, while the unstable orbits form chaotic layers which surround the islands. A key property of these islands is that the ratio $\orr/\ot$, often called the rotation number, remains constant there and forms a dynamical `plateau' \cite{cont02}.

The quadratic symmetry described above is preserved for any value of $\alpha_{22}$, however, and thus the equations of motion remain integrable. On the other hand, for non-zero values of the parameter $\alpha_{Q}$ and spin, the spacetime \eqref{eq:genmet} does not admit the Carter symmetry. Therefore, for astrophysical objects described by the line element \eqref{eq:genmet} with $\alpha_{Q} \neq 0$, Birkhoff islands form within the phase space \cite{dest20}. Depending on the exact value of the parameters $M, a, \alpha_{22}$, and $\alpha_{Q}$, the location and width of the islands vary \cite{luk,dest20}. More details on the properties of the metric \eqref{eq:genmet} and its relationship with existing astrophysical constraints can be found in Ref. \cite{dest20}.

It is the purpose of this work to explore the theoretical manifestation of a plateau at the level of the gravitational waveform. In particular, a spacetime for which the geodesic equations are non-integrable will possess a scattered series of islands. When a companion object passes through one of these islands it undergoes an abrupt orbital evolution, which leads to a sudden jump in the GW frequencies. The frequency jumps we observe are similar to that which is sometimes seen in pulsar timing experiments, namely `glitches'. For better or worse, jumps of this sort are also known to occur due to effects of instrumental origin \cite{path1,ed20}. It is important therefore that the phenomenology of non-integrability is better understood, as meaningful physics may be discarded if it is \emph{a prori} assumed that frequency jumps are not of astrophysical origin.

\noindent{\bf\emph{Orbital evolution and the kludge scheme.}} The orbital dynamics governing an EMRI are not purely geodesic since GW emission leads to a dissipation of the particle 4-momenta. In general, the equations describing the motion of a particle in a spacetime are the Einstein equations (or some appropriate generalization) together with the conversation laws $T^{\alpha \beta}_{\phantom{\alpha \beta} ;\beta} = 0$. Treating backreaction at the linear level, the conservation laws lead to the MiSaTaQuWa equations \cite{misa1,misa2}, which are known to be equivalent to the geodesic equations on a \emph{modified} spacetime $\tilde{g}_{\mu \nu} = g_{\mu \nu} + h_{\mu \nu}^{R}$, where the superscript $R$ stands for the regularized metric perturbation (see Sec. 19 of Ref. \cite{pois} for a discussion). This latter term, although necessarily small and satisfying $||h_{\mu \nu}^{R}|| \ll ||g_{\mu \nu}||$, is time-dependent and therefore allows for the particle to potentially enter and leave Birkhoff islands during its orbital lifetime.

In any case, the above shows that self-force sourced by backreaction can be modeled by introducing time-dependent shifts into the momenta $p^{\alpha}$ of the particle, which can be related to the specific energy and angular momentum through $E = - p^{t} \partial_{t}$ and $L_{z} = p^{\phi} \partial_{\phi}$. This observation forms the basis for the \emph{adiabatic} approximation introduced by Mino \cite{mino} (see also Ref. \cite{gal}), valid when the change in any quantity that characterizes the orbit (such as energy or angular momentum) is sufficiently small over a single orbit \cite{sago}. In this sense, taking an average of the MiSaTaQuWa equations essentially leads to some evolution equations for $E$ and $L_{z}$ and the remaining momenta. These equations are still, however, relatively difficult to work with. A hybrid \emph{kludge} scheme \cite{kludge1,glamp07} can instead be used where the aforementioned equations are expanded up to some desired post-Newtonian and multipolar orders. 

In this Letter, we adopt the \emph{kludge} scheme described above to perform the orbital evolutions. Up to second post-Newtonian order the relevant equations for the evolution of an orbit are lengthy, though are given explicitly by equations (37)--(39) in Ref. \cite{gair06}. These are the equations used here, though with one important modification.

In particular, at second post-Newtonian order, the \emph{kludge} scheme involves the mass quadrupole moment $M_{2}$ of the system.  To construct the non-Kerr inspiral we apply a modification to the mass quadrupole moment, similar to that in Refs. \cite{Gair:2007kr,luk}, where $M_2=-M a^2-\frac{1}{3}M^3 \alpha_Q$ for the metric \eqref{eq:genmet}. The parameter $\alpha_{22}$ introduces current multiple moments that appear at higher-PN order, and therefore does not explicitly enter into the \emph{kludge} equations of motion. While the exact definition for the multipole moments can depend on the theory under consideration \cite{papp15}, we have checked that our results are qualitatively unchanged for different choices of $M_{2}$ (e.g., $M_{2} = M_{2}^{\text{Kerr}}$). We further linearize the adiabatic energy and angular momentum evolution as in Ref. \cite{Canizares}, viz. $E_1=E_0+dE/dt|_0 N_r\, T_r$ and $L_{z,1}=L_{z,0}+ dL_z/dt|_0 N_r\, T_r$, where $E_0,\,L_{z,0}$ are the initial energy and angular momentum, respectively, $dE/dt|_0,\,dL_z/dt|_0$ are the radiation loss rates calculated  using the \emph{kludge} expressions at the beginning of the inspiral, respectively, and $T_r$ is the time that the orbit takes to travel from the periastron to apoastron and back. The aforementioned equations are updated every $N_r$ cycles for the whole EMRI evolution.

\noindent{\bf\emph{Gravitational waves and frequency evolution.}} We model the gravitational waveform using the Einstein-quadrupole approximation to understand the main phenomenology of the transient features acquired through passage of an island. A more sophisticated approach would involve solving the (appropriately generalized, see, e.g., Refs. \cite{suv19a,suv19b}) Teukolsky equations directly to deduce the GW characteristics through the Weyl scalars. However, such an analysis is considerably more complicated and does not help to elucidate the main new features presented in this work: the existence of `glitches' from chaotic inspirals. 

In general, an incoming GW can be projected onto the mutually orthogonal $+$ and $\times$ polarization states by introducing two vectors $\boldsymbol{p} = \boldsymbol{n} \times {\boldsymbol{Z}} / |\boldsymbol{n} \times {\boldsymbol{Z}}|$ and $\boldsymbol{q} = \boldsymbol{p} \times \boldsymbol{n}$, which are defined in terms of a unit vector $\boldsymbol{n}$ which points from the source to the detector for particle position $Z^i(t)$. Note that some modified theories of gravity predict the existence of up to six polarization states, a detection of which would also conclusively signal the breakdown of GR \cite{liu20}. These extra polarization states are not, however, tied to the symmetries of the spacetime, since even a Kerr black hole can emit scalar-like waves in an $f(R)$ theory, for instance \cite{suv19a}, and we assume they are negligible. At a luminosity distance $d$ from the source and in the quadrupole approximation, the GW amplitudes read $h_{+,\times}(t)=2 \mu \epsilon^{+,\times}_{ij}\left[a^i(t)Z^j(t)+v^i(t)v^j(t)\right]/d$ for velocity, $v^i(t)=dZ^i/dt$, and acceleration $a^i(t)=d^2Z^i/dt^2$ \cite{Canizares}, where the $+$ and $\times$ polarization tensors have components $\epsilon_+^{ij}=p^i p^j-q^i q^j$, and $\epsilon_\times^{ij}=p^i q^j+p^j q^i$, respectively.

The first of the deformation parameters introduced in \eqref{eq:genmet}, $\alpha_{22}$, has the strongest effect on observables when the object is near the extremal limit $a \lesssim M$ \cite{joh13}. Since accretion torques are expected to spin-up super-massive BHs that exist within galactic centers \cite{thorne74}, even a small value of $\alpha_{22}$ can lead to significant modifications in the orbital dynamics \cite{dest20}. We thus consider a near-extremal object with $a= 0.99 M$ throughout.

\begin{figure}[t]
\includegraphics[scale=0.4]{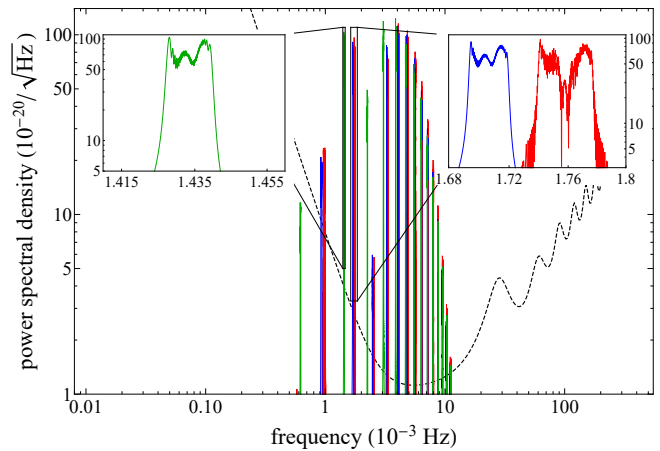}
\caption{Power spectral density of a Kerr (green), a non-Kerr (red) and a deformed-Kerr (blue) EMRI consisting of a light companion with $\mu=M_\odot$ and a supermassive compact object with $M=10^6 M_\odot$ ($\mu/M=10^{-6}$) and $a=0.99M$ at a luminosity distance $d=50 \text{ Mpc}$, over-plotted on the LISA power spectral density curve with $\text{SNR}=1$ (dashed curve). All small mass companions begin their inspiral with $E_0=0.95 \mu$, $L_{z,0}=3 M \mu$ and initial conditions $r(0)=5.70903M,\,\dot{r}(0)=0.1, \,\theta(0)=\pi/2$, with the remaining $\dot{\theta}(0)$ being defined by the constraint equation (7) in \cite{dest20}. The most prominent Fourier peaks are presented in the zoomed inlays, while the remaining peaks correspond to higher harmonics. The evolution time of the EMRI is $t=2 \times 10^6 M$, which corresponds, roughly, to $3$ months of detection. The deformation parameters are $\alpha_Q=-1.8$ for the non-Kerr and $\alpha_{22}=-2.2$ for the deformed-Kerr inspiral, while for the Kerr inspiral $\alpha_Q=\alpha_{22}=0$.}
\label{detectability}
\end{figure} 

\begin{figure*}[t]
\includegraphics[scale=0.56]{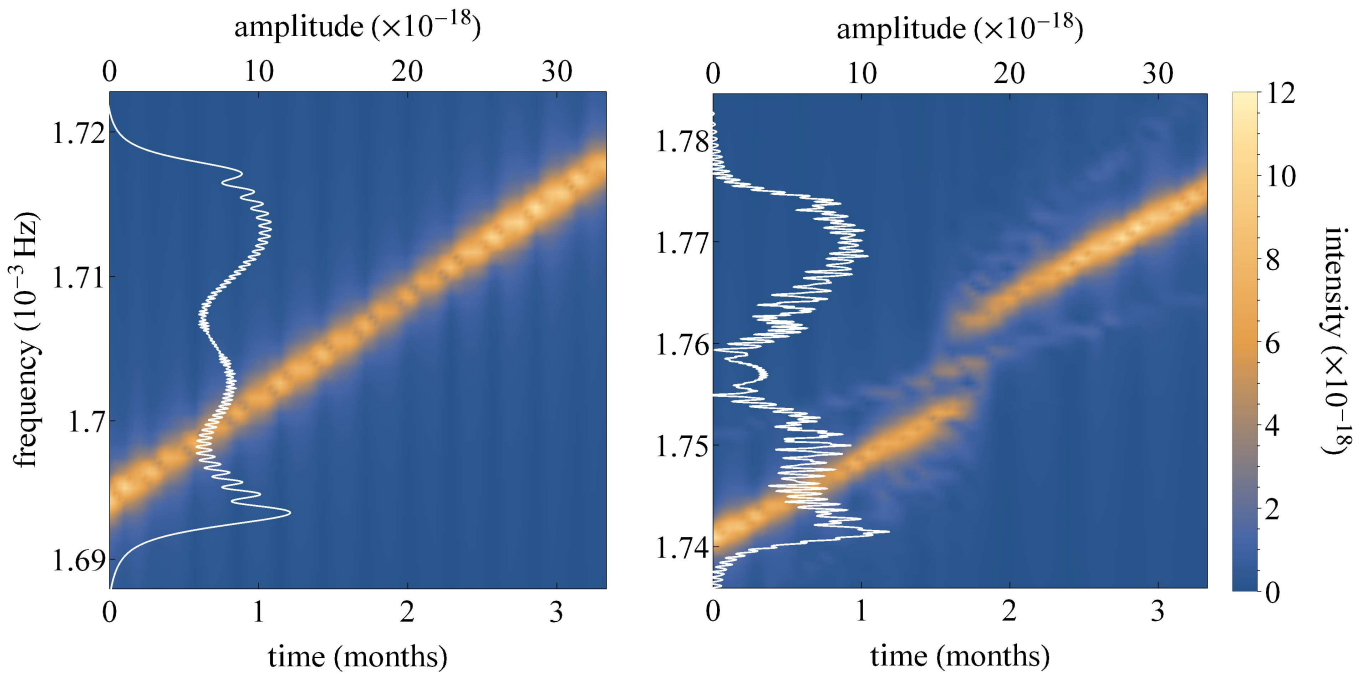}
\caption{Periodogram of a deformed-Kerr (left) and a non-Kerr EMRI (right) GW, plotted below the most prominent GW peak, as depicted in the right inlay of Fig. \ref{detectability}.}
\label{periodogram}
\end{figure*}

Although the LISA data stream consists of two linearly independent channels, with each one being better suited to the detection of particular types of signals \cite{cutler98}, we will work, for simplicity, within the single channel approximation and neglect any noise in the data stream to demonstrate the main features of detection. In what follows, we compare frequency evolution and power spectral densities (PSDs) of two EMRIs, consisting of a small-mass companion (a stellar mass BH or a neutron star) with $\mu=M_\odot$ and a supermassive compact object with $M=10^6 M_\odot$ ($\mu/M=10^{-6}$); the first evolves on a non-Kerr spacetime and crosses a Birkhoff island of $2/3$-resonance, while the second evolves on a deformed-Kerr spacetime and passes through a $2/3$-resonance, though no Birkhoff islands are present anywhere in the phase space. The detectability of the PSD of such EMRIs is demonstrated in Fig. \ref{detectability}, where we assume a source at $d=50 \text{ Mpc}$, where the LISA PSD is also plotted with signal-to-noise ratio (SNR) equal to unity. For completeness, we further include the PSD of a Kerr EMRI. The most prominent peaks, including various higher harmonics, appear in the range of maximal sensitivity of LISA, between $10^{-3}-10^{-2}$ Hz. Although the PSD peaks of Kerr and deformed-Kerr EMRIs, which share the same fundamental spacetime symmetries, have smooth profiles (see inlays in Fig. \ref{detectability}), non-Kerr EMRIs, which cross Birkhoff islands, will typically undergo a frequency modulation during the crossing, where the PSD peaks (and the amplitude of the Fourier spectrum of the GW) abruptly decrease by up to $2$ orders of magnitude (see right inlay in Fig. \ref{detectability}). This substantial drop in amplitude should lead to a significant effect in the frequency evolution of a non-Kerr EMRI.

In Fig. \ref{periodogram} we show the frequency evolution of the most prominent spectral peaks from a deformed-Kerr and a non-Kerr EMRI with the same initial conditions (we have chosen the deformation parameters for each EMRI so that the initial rotation numbers of the orbits agree to $\sim 0.1\%$). When the underlying spacetime symmetry is similar to that of Kerr, the frequency evolution follows a linear growth as the object inspirals towards its supermassive companion. Although the deformed-Kerr EMRI passes through a $2/3$-resonance, no obvious effect is present. On the other hand, when a non-Kerr EMRI crosses through a Birkhoff island of $2/3$-resonance a `glitch' interrupts the linear evolution of the GW frequencies. Although we only demonstrate the effect of the most prominent resonant island (corresponding to a width of $\sim 0.05M$ and average crossing time $\sim 4 \times 10^4 M\sim 3$ days for the parameters used), other islands surrounding less prominent resonant stable orbits still exhibit a similar, but less profound, glitch. 

\noindent{\bf\emph{Discussion.}} Future space-borne GW detectors, such as LISA and Taiji \cite{lisafreq,taiji}, will unlock the detection realm to a wider range of GW sources. The detection of GWs from EMRIs, which consist of small mass companions orbiting around supermassive compact objects, in much wider orbits than those already observed by ground-based detectors, will provide significant information on the validity of GR in the strong-field region. If the astrophysical environment around EMRIs `modifies' the underlying theory of gravity in such a way that the integrability of the equations of motion is broken, then the orbital phase space of the small companion will contain a series of Birkhoff islands, by virtue of the chaotic dynamics \cite{arnold89}. 

By employing the hybrid \emph{kludge} scheme \cite{kludge1,glamp07,gair06} to evolve EMRIs in a deformed and a non-Kerr spacetime, introduced in \cite{dest20}, and the Einstein-quadrupole approximation \cite{thorne80} to model GW emission, we have explored the detectability of transient phenomena in the gravitational waveform, which designate a crossing through a Birkhoff island, and thus probe spacetime symmetry and constrain a potential departure from a Kerr description.

Our results indicate that the continuous evolution of GW frequencies is abruptly, though consistently, broken during a Birkhoff island crossing, which leads to a glitch, similar to that seen in pulsars and in the LISA pathfinder data stream, which until now is assumed to be instrumental noise. The glitches displayed here are present in the majority of the periodograms of the prominent GW frequencies and have a clear astrophysical origin. 

Taking into account that a plethora of initial orbital parameters can eventually lead an inspiral through an island of strong resonance \cite{apo09,luk}, and that an object can potentially occupy an island for up to a week \cite{Lukes-Gerakopoulos:2014dpa}, we speculate that space-borne detectors should be able to unveil or constrain the existence of chaotic phenomena in EMRIs, which may be associated with deviations from GR (though cf. Refs. \cite{bara07,amaro11,kiu04,spin1}). Furthermore, we argue that such abrupt frequency jumps may have a true astrophysical origin, thus discarding them \emph{a priori} from the data stream as instrumental artifacts may miss potential `smoking gun' physics.

\begin{acknowledgments}
\noindent{\bf\emph{Acknowledgments.}} The authors would like to thank Georgios Lukes-Gerakopoulos for helpful discussions. AGS is supported by the Alexander von Humboldt Foundation.
\end{acknowledgments}

\end{document}